\documentstyle[preprint,aps,epsfig]{revtex}
\begin{document}
\draft
\title{Main phenomenological features of the \\ spontaneous CP
violation in $SU(3)_L \otimes U(1)_Y$ models}
\author { D. G\'omez Dumm\footnote{Fellow of CONICET (Argentina)} \\
Laboratorio de F\'\i sica Te\'orica, \\
Departamento de F\'\i sica, U.N.L.P., \\
c.c. 67, 1900, La Plata, Argentina\footnote{e-mail:
dumm@venus.fisica.unlp.edu.ar}}

\maketitle

\begin{abstract}
We analyze the phenomenological consequences of assuming spontaneous CP
violation in an $SU(3)_L \otimes U(1)_Y$ model with three Higgs triplets
and one sextuplet. After the identification of the relevant physical
scalars, we estimate the contributions to the parameters $\Delta m_K$,
$\varepsilon$  and $\varepsilon'$ coming from the Higgs-fermion couplings.
\end{abstract}

\pacs{ }

\noindent \underline{{\large Introduction}}

Since the discovery of CP violation in 1964, the understanding of its
origin has represented a very attractive problem from the theoretical
point of view.  Within the context of the Standard Model (SM), the present
experimental situation can be successfully explained through the
well-known Kobayashi$-$Maskawa mechanism. This requires explicit CP
nonconservation at the Lagrangian level.  However, as it was firstly
pointed out by T.\ D.\ Lee \cite{lee}, the CP symmetry can also be
spontaneously broken if the SM Higgs sector is extended.

In this work, we consider an extension of the SM based on the $SU(3)_C
\otimes SU(3)_{L}\otimes U(1)_{Y}$ gauge symmetry group, which has been
recently proposed by Pisano and Pleitez \cite{vic} and by Frampton
\cite{fra}. The fermions are here organized into $SU(3)_L$ triplets and
antitriplets, in such a way that the model is anomaly-free only when the
number of lepton families is a multiple of the number of colors.  The
$SU(3)_L$ structure leads to the introduction of new unobserved particles,
such as exotic ``leptoquarks'' and new vector bosons of double, single and
zero electric charge. In addition, a large number of scalar fields is
required in order to provide the fermion masses.

\renewcommand{\thefootnote}{a}

In a previous paper \cite{cp}, we studied the possibility of finding
spontaneous CP violation within this scenario\footnote{The problem of CP
violation in this kind of model has been also treated in Ref.\
\cite{liung}, in connection with lepton-number-violating interactions.}.
By analyzing the Higgs potential, we have shown that the mechanism can be
triggered for a certain range of the involved parameters. This could give
rise to measurable effects through the presence of complex couplings in
the Yukawa Lagrangian.

The purpose of this article is to analyze the phenomenology originated by
the nontrivial structure of the neutral and single charged Higgs sector of
the model. We study the effects of scalar-mediated flavor-changing neutral
interactions (FCNI's), as well as the possibility of spontaneous CP
violation.  In particular, we show that this scheme is capable of
explaining the observed CP violation phenomena related to kaon decays.

We proceed first by identifying the scalar mass eigenstates when CP is
conserved and then when CP is spontaneously broken. Afterwards we compute
the contributions to the mass difference $m_{K_L}-m_{K_S}$ and the
CP-violating parameters $\varepsilon$ and $\varepsilon'$. Finally, we
discuss the compatibility of the model with the experimental results.

\hfill

\noindent \underline{{\large Fermions, scalars and gauge bosons}}

As was mentioned above, the left-handed fermions appear organized into
$SU(3)_{L}$ triplets. These are
\begin{displaymath}
\Psi_{lL} = \left( \begin{array}{l} \nu_{l} \\ l \\ l^{c} \end{array}
 \right)_{L} \sim (\underline{3}, 0)
\hspace{2cm}
Q_{1L} = \left( \begin{array}{l} u_{1} \\ d_{1} \\ J_{1} \end{array}
 \right)_{L} \sim (\underline{3}, \frac{2}{3})
\end{displaymath}
\begin{displaymath}
Q_{2L} = \left( \begin{array}{l} J_{2} \\ u_{2} \\ d_{2} \end{array}
 \right)_{L} \rule{0cm}{2cm} , \;
Q_{3L} = \left( \begin{array}{l} J_{3} \\ u_{3} \\ d_{3} \end{array}
 \right)_{L} \sim (\underline{3}^{\ast}, -\frac{1}{3})
\end{displaymath}
where $l=e,\mu,\tau$. The first and second entries following each triplet
denote the corresponding $SU(3)_L$ representation and $Y$ quantum number
respectively. Here, in order to complete the quark triplets, the inclusion
of ``leptoquarks'' of charge $\frac{5}{3}$ and $-\frac{4}{3}$ has been
necessary. Notice that two of the quark families belong to the
$\underline{3}$ representation, while the third one transforms as a
$\underline{3}^\ast$. This gives rise to FCNI's at the tree level.

The right-handed fields are introduced in the model as $SU(3)_{L}$
singlets
\[
u_{iR} \sim (\underline{1},\frac{2}{3}) \;\;\;
d_{iR} \sim (\underline{1},-\frac{1}{3}) \;\;\;
J_{1R} \sim (\underline{1},\frac{5}{3}) \;\;\;
J_{2,3R} \sim (\underline{1},-\frac{4}{3}) \;\;\;
l_{R} \sim (\underline{1},-1)
\]
In both cases, the values of $Y$ verify the Gell-Mann$-$Nishijima relation
\[ Q = Y + T_{3} - \sqrt{3}\, T_{8} \]

The fermions acquire masses through Yukawa-like couplings with Higgs
bosons. These scalar fields are arranged into three $SU(3)_{L}$ triplets
with $Y$ values 1, 0 and -1 respectively,
\begin{displaymath}
\begin{array}{lll}
\rho = \left( \begin{array}{l} \rho^{+} \\ \rho^{0} \\ \rho^{++}
\end{array} \right) \, ,\; &
\eta = \left( \begin{array}{l} \eta^{0} \\ \eta_{1}^{-} \\
\eta_{2}^{+} \end{array} \right) \, ,\, &
\chi = \left( \begin{array}{l} \chi^{-} \\ \chi^{--} \\ \chi^{0}
\end{array} \right)
\end{array}
\end{displaymath}
and one $Y\!=\!0$ sextuplet
\begin{displaymath}
S = \left( \begin{array}{lll} \sigma_{1}^{0} & h_{2}^{-} & h_{1}^{+} \\
h_{2}^{-} & H_{1}^{--} & \sigma_{2}^{0} \\
h_{1}^{+} & \sigma_{2}^{0} & H_{2}^{++} \end{array} \right)
\end{displaymath}
Finally, the model also contains new charged vector bosons, namely $V^+$
and $U^{++}$, which together with their charge conjugated and a new
neutral vector field $Z'^0$ complete the $SU(3)_L$ gauge octet.

\hfill

\noindent \underline{{\large Higgs potential and CP violation}}

In order to yield the expected spontaneous gauge symmetry breakdown, the
scalar potential should be minimized when the scalar fields take the
values
\begin{equation}
\begin{array}{llll}
\langle \rho \rangle = \left( \begin{array}{c} 0 \\
v_{\rho} e^{i\theta_\rho} \\ 0 \end{array} \right) \, , &
\langle \eta \rangle = \left( \begin{array}{c} v_{\eta} e^{i\theta_\eta}
\\ 0 \\ 0 \end{array} \right) \, , &
\langle \chi \rangle = \left( \begin{array}{c} 0 \\ 0 \\ v_{\chi}
e^{i\theta_\chi} \end{array} \right) \, , &
\langle S \rangle = \left( \begin{array}{ccc} \; 0\; & 0 & 0 \\ 0 & 0 & v_{s}
e^{i\theta_s} \\ 0 & v_{s} e^{i\theta_s} & 0 \end{array} \right)
\end{array}
\label{vevs}
\end{equation}
where we have explicitly written the nonzero VEV's as complex numbers. If
the gauge symmetry of ${\cal L}$ is now taken into account, the phases
$\theta_\chi$ and $\theta_\rho$ can be chosen to be equal to zero without
any loss of generality. The remaining phases $\theta_\eta$ and $\theta_s$
survive as CP-violating parameters in the final Lagrangian of the theory.

In (\ref{vevs}) we have chosen $\langle \sigma_{1}^{0} \rangle = 0$,
preventing in this way the presence of neutrino mass terms. It can be
shown, however, that if $\langle \sigma_{1}^{0} \rangle$ has to be kept
equal to zero, a fine tuning between some of the coupling constants in the
Higgs potential $V(\eta ,\rho ,\chi ,S)$ \cite{cp,foot} becomes necessary.
One way of avoiding this fine tuning and achieving the above vacuum
expectation values is through the introduction of a particular set of
discrete symmetries \cite{foot}:
\begin{equation}
\begin{array}{rclcrcl}
Q_{1L} & \rightarrow & - Q_{1L} &
\hspace{.1cm} & \eta & \rightarrow & - \eta \\
Q_{jL} & \rightarrow & - i Q_{jL} &
\hspace{.1cm} & \rho, \, \chi  & \rightarrow & i \; \rho, \, \chi \\
\Psi_{lL} & \rightarrow & i \Psi_{lL} &
\hspace{.1cm} & S & \rightarrow & - S \\
u_{jR} & \rightarrow & u_{jR} &
\hspace{.1cm} & J_{1R} & \rightarrow & i J_{1R} \\
d_{jR} & \rightarrow & i d_{jR} &
\hspace{.1cm} & J_{2,3R} & \rightarrow & J_{2,3R}
\label{simetrias}
\end{array}
\end{equation}
where $j=1,2,3$ and $l=e,\mu,\tau$. At the same time, the presence of
these symmetries ensures that the lepton and baryon numbers result
conserved in the final Lagrangian separately.

Once we have obtained the desired nonzero VEV's in (\ref{vevs}), we
demand them to satisfy the relations
\begin{equation}
v_{\chi}^{2} \gg v_{\eta}^{2} \approx v_{\rho}^{2} \approx v_{s}^{2}
\label{rango}
\end{equation}
in order to guarantee that the gauge symmetry breakdown follows the
hierarchy $SU(3)_{L} \otimes U(1)_{Y} \rightarrow SU(2)_{L} \otimes
U(1)_{Y} \rightarrow U(1)_{em}$. In addition, from the Higgs kinetic term
in ${\cal L}$ these VEV's are related by
\begin{equation}
\frac{g}{2} \left( v_\eta^2+v_\rho^2+\frac{v_s^2}{2}\right) =m_W^2
\label{mw}
\end{equation}

\hfill

Let us now turn to the phenomenology of the model, in particular to those
effects coming from the enlarged scalar sector. To this end, we first need
to isolate the physical scalars of the theory, which result from the
diagonalization of the Higgs mass matrices. Notice that the latter receive
contributions from terms which are proportional to $(\varphi_i^\dagger
\varphi_i)$, $(\varphi_i^\dagger\varphi_i) (\varphi_j^\dagger \varphi_j)$,
and others involving the product $S^\dagger S$, $\varphi_i$ being any of
the Higgs triplets and $S$ the Higgs sextuplet.  Considering the
symmetries in (\ref{simetrias}), the remaining contributions come from the
non-self-Hermitian terms present in the potential,
\begin{eqnarray}
& & f_{1} \, \mu_1\,\epsilon_{ijk} \,\eta_{i} \rho_{j} \chi_{k} +
f_{2} \,\mu_2\,\chi^{\dag} S \rho^{\ast} + f_{3} \, \epsilon_{ijk}
\, \eta_{i} \rho_{j} (S \rho^{\ast})_{k} \nonumber \\
& & + f_{4} \, \epsilon_{ijk} \,\eta_{i} (S \chi^{\ast})_{j} \chi_{k} +
f_{5} \, \epsilon_{ijk} \, \epsilon_{lmn} \, S_{il} S_{jm} \eta_{k} \eta_{n}
+ \mbox{ h.c.}
\end{eqnarray}
where the parameters $\mu_{1,2}$ have dimensions of mass. As is shown in
Ref. \cite{cp}, this non-self-Hermitian piece of ${\cal L}$ yields
spontaneous CP violation for an appropriate range of the parameters $f_i$
and $\mu_i$.

The main contributions to $\Delta m_K$ and $\varepsilon$ are expected to
come from the neutral flavor changing interactions, therefore we start with
the analysis of the neutral scalar mass matrix. Due to the symmetries
(\ref{simetrias}), the fields Re($\sigma_1^0$) and Im($\sigma_1^0$) decouple
to the rest, becoming exact mass eigenstates. The other neutral scalar
fields, namely the real and imaginary parts
of $\eta^0$, $\rho^0$, $\chi^0$ and $\sigma_2^0$, result mixed by an $8\times
8$ real mass matrix. As a first step, we identify in this sector two $m=0$
eigenstates, the neutral scalar Goldstone bosons. These are given by
\begin{eqnarray}
G_1^0 & = & \frac{1}{\beta}\left[ v_\rho\,\mbox{Im}(\rho^0) -
v_\eta\,\mbox{Im} (\eta^0 e^{-i\theta_\eta}) + v_s\,\mbox{Im}(\sigma^0_2
e^{-i\theta_s})\right] \nonumber \\
G_2^0 & = & \frac{1}{\beta N^2}\left[ - v_\rho \alpha^2\,\mbox{Im}(\rho^0)
- v_\rho^2 v_\eta\,\mbox{Im}(\eta^0 e^{-i\theta_\eta}) + v_\rho^2 v_s\,
\mbox{Im}(\sigma^0_2 e^{-i\theta_s})+\beta^2 v_x\,\mbox{Im}(\chi^0)\right]
\label{gold}
\end{eqnarray}
where we have defined $\alpha=(v_\eta^2 + v_s^2)^\frac{1}{2}$, $\beta=
(\alpha^2 + v_\rho^2)^\frac{1}{2}$, $N^2=(\alpha^2 v_\rho^2 + \beta^2
v_\chi^2)^\frac{1}{2}$.

The identification of the massive scalar fields is now necessary. This is
not a trivial task, in view of the many terms which are present in the
Higgs potential. However, it is possible to get more insight into the
physics of this sector if one considers some proper assumptions.

Let us present the potential in such a way that the terms of the form
$(\varphi_i^\dagger \varphi_i)(\varphi_j^\dagger \varphi_j)$, appear
written as $\lambda_{ij} (\varphi^\dagger_i \varphi_i - |v_i|^2 +
\varphi^\dagger_j \varphi_j - |v_j|^2)^2$, and analogously for the terms
involving $S^\dagger S$ \cite{foot}. The basic assumption will be to
consider that no further scales are introduced by the parameters in the
Higgs potential once it has been written in this ``complete squared'' way.
That is, we take all the $\lambda_{ij}$ parameters, together with the
$f_i$ ones, to be of the same order of magnitude, say ${\cal O}(\lambda)$.
Then, taking into account the relation (\ref{rango}), we will suppose that
the ``massive'' parameters $\mu_{1,2}$ lay in the range $v_\eta < \mu_i <
v_\chi$, so that no other mass scale is needed.

Next, we change to the new basis
\begin{eqnarray}
\phi_1 & = & \frac{1}{\beta}\left[ v_\eta\,\mbox{Re}(\eta^0
e^{-i\theta_\eta}) + v_\rho\,\mbox{Re}(\rho^0) + v_s\,\mbox{Re}(\sigma^0_2
e^{-i\theta_s})\right] \nonumber \\
\phi_2 & = & \mbox{Re}(\chi^0) \nonumber \\
\phi_3 & = & \frac{1}{\beta\alpha}\left[- v_\eta v_\rho\,\mbox{Re}(\eta^0
e^{-i\theta_\eta}) + \alpha^2 \,\mbox{Re}(\rho^0) - v_s v_\rho\,
\mbox{Re}(\sigma^0_2 e^{-i\theta_s})\right]
\label{base1} \\
\phi_4 & = & \frac{1}{\alpha N^2}\left[ v_\eta v_\rho v_\chi \,
\mbox{Im}(\eta^0 e^{-i\theta_\eta}) + v_\chi \alpha^2\,\mbox{Im}(\rho^0) -
v_s v_\rho v_\chi\,\mbox{Im}(\sigma^0_2 e^{-i\theta_s}) +
v_\rho \alpha^2\,\mbox{Im}(\chi^0) \right] \nonumber \\
\phi_5 & = & \frac{1}{\alpha}\left[ -v_s \,\mbox{Re}(\eta^0
e^{-i\theta_\eta}) + v_\eta \,\mbox{Re}(\sigma_2^0 e^{-i\theta_s})\right]
\nonumber \\
\phi_6 & = & \frac{1}{\alpha}\left[ v_s\,\mbox{Im}(\eta^0
e^{-i\theta_\eta}) +  v_\eta \,\mbox{Im}(\sigma_2^0
e^{-i\theta_s})\right]\nonumber
\end{eqnarray}
It can be seen that these will be the approximate mass eigenstates when
$v_\eta\ll\mu\ll v_\chi$. The respective masses will follow in this case
the hierarchy
\begin{equation}
\begin{array}{rcl}
m_{\phi_1}^2 & \sim & \lambda v_\eta^2 \nonumber \\
m_{\phi_4}^2,m_{\phi_3}^2 & \sim & \lambda \mu v_\chi \nonumber \\
\; m_{\phi_5}^2, m_{\phi_6}^2, m_{\phi_2}^2\; & \sim & \lambda v_\chi^2
\label{masas1}
\end{array}
\end{equation}
Setting $\theta_\eta=\theta_s=0$, we see that $\phi_4$ and $\phi_6$ have
to be CP-odd, while the other scalar fields are CP even.

Let us now examine the quark-Higgs Yukawa couplings of the Lagrangian. In
terms of the $SU(3)_L$ multiplets components, we have for the neutral
Higgs sector
\begin{eqnarray}
-{\cal L}_Y&=&\frac{1}{v_{\rho}^{*}}\rho^{0*}\bar U_L\hat M^uU_R+
\frac{1}{v_{\eta}}\left(\eta^0-\frac{v_{\eta}}{v_{\rho}^{*}}
\rho^{0*}\right)\bar U_L\Delta^u\hat M^uU_R\nonumber \\ & &\mbox{}
+\frac{1}{v_{\eta}^{*}}\eta^{0*}\bar D_L\hat M^dD_R+\frac{1}{v_{\rho}}
\left(\rho^0-\frac{v_{\rho}}{v_{\eta}^{*}}
\eta^{0*} \right)\bar D_L\Delta^d\hat M^dD_R
+ \mbox{ h.c.}
\label{yuka}
\end{eqnarray}
where $\hat M^u$ and $\hat M^d$ represent the diagonalized quark mass
matrices, and we have also defined $\Delta^{u,d}\equiv V_L^{u,d\dagger}
\mbox{diag} (1,0,0) V_L^{u,d}$. The unitary matrices $V_L^{u,d}$, used for
changing to the quark mass eigenbasis, are related by
\begin{equation}
V_{L}^{u\dagger} V_{L}^d = V_{CKM}
\label{vkm}
\end{equation}
{}From (\ref{yuka}), the neutral Higgs sector is nondiagonal in flavor. We
expect then to find significant contributions to $\Delta m_K$ and
$\varepsilon$ coming from tree level FCNI's.

If we choose now the basis defined in (\ref{base1}) for the scalar fields,
the Yukawa couplings take the form
\begin{eqnarray}
{\cal L}_Y & = & \frac{1}{\beta} (\bar U\hat M^u U) \phi_1 + \left[
\frac{\alpha}{\beta v_\rho}\bar U \hat M^u U - \frac{\beta}{\alpha v_\rho}
\bar U (A^u_+ + A^u_- \gamma_5) U\right] \phi_3 \nonumber \\
& & + i\,\left[-\frac{\alpha}{\beta v_\rho}\bar U \hat M^u \gamma_5 U +
\frac{\beta}{\alpha v_\rho} \bar U (A^u_- + A^u_+ \gamma_5) U\right]
\phi_4 + ( u \leftrightarrow d)
\label{lagbase1}
\end{eqnarray}
where $A^u_{\pm}=\Delta^u \hat M^u\pm \hat M^u\Delta^u$. From this
expression, the identification of the $\phi_1$ as the ``standard'' flavor
conserving Higgs is immediate. The lowest (tree level) correction to
$\Delta m_K$ is given by the scalars $\phi_4$ and $\phi_3$, which have
intermediate masses (see Eq.\ (\ref{masas1})). On the other hand,
according to the CP parity for the involved neutral fields, ${\cal L}_Y$
is found to be CP invariant, not depending on $\theta_\eta$ and
$\theta_s$. In this way, the CP violation effects will be due to the
mixing among the scalars, hence they will be suppressed at least by a mass
scale ratio such as $\mu /v_\chi$ or $v_\eta /\mu$.

\hfill

We will look into the following for the possibility of finding spontaneous
CP breakdown. We have shown in Ref.\ \cite{cp} that this can be achieved
only if the mass parameters $\mu_i$ are close to the VEV $v_\chi$. This
enables the presence of just two relevant mass scales in the Higgs sector,
say $v_\chi$ and $v_\eta$. The Higgs potential will be minimized in this
case by nonzero values of $\theta_\eta$ and $\theta_s$. These values can
be exactly calculated in terms of $v_\eta$, $v_\rho$, $v_\chi$, $v_s$ and
the potential parameters $f_i$ and $\mu_i$ \cite{cp}.

Once the relation $\mu\sim v_\chi$ is assumed, the fields defined in
(\ref{base1}) are no longer approximate mass eigenstates. In order to find
the exact physical particles, let us once again change to a new basis,
namely
\begin{eqnarray}
{h^0_2}' & = & \frac{1}{N_2^2}\,[-c \phi_3 + b \phi_4 + d \phi_6]
\nonumber \\
{h^0_3}' & = & \frac{1}{N_4^2}\,[-d \phi_3 - c \frac{\beta v_\chi}{N^2}
\phi_5 + b \phi_6] \nonumber \\
{H^0_2}' & = & \frac{1}{N_1^2}\,[a \phi_3 + b \phi_4 + d \phi_6] \\
{H^0_3}' & = & \frac{1}{N_3^2}\,[-d \phi_3 + a \frac{\beta v_\chi}{N^2}
\phi_5 + b \phi_6] \nonumber \\
{h^0_1}' & = & \phi_1 \nonumber \\
{H^0_1}' & = & \phi_2 \nonumber
\end{eqnarray}
where the $N_i^{-2}$ are normalization factors, verifying $N_1\simeq N_3$
and $N_2\simeq N_4$. The parameters $a$, $b$, $c$ and $d$ are functions of
the coupling constants $f_i$ and $\mu_i$:
\begin{eqnarray}
a & = & \frac{2\beta^2 v_\chi}{v_\rho \alpha^2} (f_1 \mu_1 v_\eta
\cos\theta_\eta + f_2 \mu_2 v_s\cos\theta_s) \nonumber \\
b & = & \frac{2\beta v_\chi}{\alpha^2} (f_1 \mu_1 v_s\cos\theta_\eta -
f_2 \mu_2 v_\eta\cos\theta_s) \label{abcd} \\
c & = & f_4\frac{2\alpha^2 v_\chi^2}{v_\eta v_s}\cos(\theta_\eta + \theta_s)
+ \frac{2 v_\chi v_\rho}{\alpha^2} (f_1 \mu_1 \frac{v_s^2}{v_\eta}
\cos\theta_\eta + f_2 \mu_2 \frac{v_\eta^2}{v_s}\cos\theta_s) \nonumber \\
d & = & f_1 \mu_1 \frac{2 v_\chi \beta}{v_s} \sin\theta_\eta \nonumber
\end{eqnarray}
The CP violation is carried by the $\sin \theta_\eta$ in $d$.

It is now worth regarding the appearance of the mass matrix. Actually, it
is found that it blocks out into two parts, which correspond to three
$m^2\sim\lambda v_\eta^2$ fields and three $m^2\sim \lambda v_\chi^2$
fields, combinations of the ${h^0_i}'$ and the ${H^0_i}'$ ones
respectively. If we ignore the $(v_\eta^2/v_\chi^2)$ corrections due to
the ${h^0_i}'-{H^0_i}'$ mixing, we obtain finally the physical particles
\begin{equation}
h_i^0={\cal U}_{ij} {h^0_j}' \;\;\;\; H_i^0={\cal V}_{ij} {H^0_j}'
\end{equation}
whose masses remain of order $\lambda^{\frac{1}{2}} v_\eta$ and
$\lambda^{\frac{1}{2}} v_\chi$ respectively. When the Lagrangian
(\ref{lagbase1}) is written in terms of the approximate mass eigenstates
$h_i^0$ and $H_i^0$ defined above, one can immediately see that all the
resulting neutral couplings are in general nondiagonal in flavor.

\hfill

We focus our attention on the single charged scalars now. Due to the
symmetries (\ref{simetrias}), these separate into two groups, the
$\eta_{1}^{+}, \rho^{+}, h_{1}^{+}$ and the $\eta_{2}^{+}, \chi^{+},
h_{2}^{+}$ fields. The second group does not interact with the ordinary
quarks at the tree level and will not be considered here.

Following a similar procedure as the one used for the neutral scalar
sector, we change to a new basis containing the Goldstone boson:
\begin{eqnarray}
G^+ & = & \frac{1}{\beta} [-v_\eta e^{i\theta_\eta} \eta_1^+ +
v_\rho \rho^+ + v_s e^{-i\theta_s} h_1^+] \nonumber \\
V_1^+ & = & \frac{1}{\alpha} [v_s e^{i\theta_\eta} \eta_1^+ +
v_\eta e^{-i\theta_s} h_1^+] \\
V_2^+ & = & \frac{1}{\alpha\beta} [v_\eta v_\rho e^{i\theta_\eta} \eta_1^+
+ \alpha^2 \rho^+ - v_s v_\rho e^{-i\theta_s} h_1^+] \nonumber
\end{eqnarray}
where the complex $3\times 3$ mass matrix takes the form
\begin{equation}
M^2=\frac{1}{2}\left(\begin{array}{ccc} 0 & 0 & 0 \\
0 & A & B \\ 0 & B^{\ast} & C \end{array} \right)
\label{matriz}
\end{equation}
Up to the order $(v_\eta^2/v_\chi^2)$, we find for the elements $A$, $B$
and $C$ in (\ref{matriz}) the values
\begin{eqnarray}
A & = & f_1 \mu_1 v_\rho v_\chi \left( \frac{v_\eta
\cos\theta_\eta}{\alpha^2} - \frac{1}{v_\eta \cos\theta_\eta} \right)
\nonumber \\
& & + f_2 \mu_2 v_\rho v_\chi \left( \frac{v_s \cos\theta_s}{\alpha^2} -
\frac{1}{v_s \cos\theta_s} \right) -
f_4 v_\chi^2 \left( \frac{v_s \cos\theta_s}{v_\eta \cos\theta_\eta} +
 \frac{v_\eta \cos\theta_\eta}{v_s \cos\theta_s} \right)
\nonumber \\
B & = & \frac{\beta v_\chi}{\alpha^2} (-f_1\mu_1 v_s e^{-i\theta_\eta} +
f_2\mu_2 v_\eta e^{i\theta_s}) \hspace{2cm} C = \frac{|B|^2}{A}
\end{eqnarray}
Here, the CP violation is produced by the imaginary part of $B$. The
relevant $2\times 2$ complex submatrix in (\ref{matriz}) can be easily
diagonalized by means of
\begin{equation}
{\cal V}^{h^+}=\left( \begin{array}{cc} \cos\gamma &
-\sin\gamma e^{i\varphi} \\
\sin\gamma e^{-i\varphi} & \cos\gamma \end{array} \right)
\label{higgmat}
\end{equation}
where $\tan 2\gamma=2|B|/(C-A)$ and $\tan\varphi=\mbox{Im}B/\mbox{Re}B$.
Thus, to this order, the physical charged scalars are given by
\begin{equation}
\left( \begin{array}{c} h^+ \\ H^+ \end{array} \right)
= {\cal V}^{h^+}
\left( \begin{array}{c} V_1^+ \\ V_2^+ \end{array} \right)
\end{equation}
The squared mass of $H^+$ is order $\lambda v_\chi^2$, while the other
eigenvalue is zero. Going to the next order in $v_\eta^2/v_\chi^2$, the
zero is not maintained, so we have $m_{h^+}^2\sim \lambda v_\eta^2$.
Notice that the parameter $\varphi$ in ${\cal V}^{h^+}$ is a CP violation
source that will appear in the Yukawa couplings.

\hfill

\noindent \underline{{\large $SU(3)_L \otimes U(1)_Y$ effects on $\Delta
m_K$, $\varepsilon$ and $\varepsilon'$}}

The model under consideration was shown to present flavor changing neutral
currents in the scalar sector, as well as Higgs-mediated interactions with
CP violation. Hence, it is well possible to find nonnegligible
contributions to the mass difference $m_{K_L}- m_{K_S}$ and the
CP-violating parameters $\varepsilon$ and $\varepsilon'$.

First of all, we consider the tree level contribution to $\Delta m_K$
shown in Fig.\ 1, which has no analogous within the Standard Model. In the
$SU(3)_L \otimes U(1)_Y$, the diagram arose since the couplings between
the quarks and the physical scalars $h_i^0$ are nondiagonal in flavor.
After the evaluation of the effective $\Delta S=2$ Hamiltonian, and using
$\Delta m_K=2\,\mbox{Re}\langle K^0|{\cal H}_{eff} |\bar K^0 \rangle$ we
obtain the value
\begin{eqnarray}
(\Delta m_K)_{h^0} & \simeq & (\Delta_{12}^d)^2
\frac{4 f_K^2 m_K^3}{3\alpha^2 N_2^2} B_K \sum_{i=1}^3
\frac{1}{m_{h_i^0}^2} \left[ K_1 K_2 {\cal U}_{2i} {\cal U}_{3i}
\right. \nonumber \\
& & \left.+\frac{1}{4} (K_1^2 - K_2^2) ({\cal U}_{2i}^2 -
{\cal U}_{3i}^2 ) - \frac{m_d}{2 m_s} (K_1^2 + K_2^2) ({\cal U}_{2i}^2
+ {\cal U}_{3i}^2 ) \right]
\label{dm0}
\end{eqnarray}
Here, and in the following, we use the definitions
\begin{eqnarray*}
K_1 & = & \frac{\beta}{v_\rho} c - \frac{v_s}{v_\eta} b \\
K_2 & = & \frac{v_s}{v_\eta} d \\
K_3 & = & \frac{\alpha^2}{\beta v_\rho} c
\end{eqnarray*}
with $a$, $b$, $c$ and $d$ as in (\ref{abcd}). As is usually done, we used
the vacuum insertion approximation to estimate the hadronic matrix
element. The uncertainty is absorbed in the parameter $B_K$. It will be
shown below that the value in (\ref{dm0}) represents a severe constraint
for the $SU(3)_L \otimes U(1)_Y$ model Higgs sector in connection with the
flavor-changing effects.

Let us now analyze the model predictions for the $\varepsilon$ and
$\varepsilon'$ parameters. According to standard definitions, we have
\begin{mathletters}
\begin{equation}
\varepsilon = \frac{\mbox{Im} M_{12}}{\sqrt{2} \Delta m_K} +
\frac{1}{\sqrt{2}} \frac{\mbox{Im}A_0}{\mbox{Re}A_0}
\label{e}
\end{equation}
\begin{equation}
\varepsilon' = i \frac{e^{i(\delta_2-\delta_0)}}{\sqrt{2}}
\left( \frac{\mbox{Im}A_2}{\mbox{Re}A_0} -
\frac{\mbox{Im}A_0\mbox{Re}A_2}{(\mbox{Re}A_0)^2} \right)
\label{e'}
\end{equation}
\end{mathletters}
where $M_{12}=\langle K^0|{\cal H}_{eff}|\bar K^0 \rangle$ and $A_I$ is
the $K^0$ decay amplitude to the final $\pi\pi$ state with isospin $I$.
The $\delta_I$ stand for the corresponding phase shifts.

The lowest order neutral Higgs contribution to $\varepsilon$ is obtained
by taking the imaginary part of the diagram in Fig.\ 1. We get
\begin{equation}
(\mbox{Im} M_{12})_{h^0}= (\Delta_{12}^d)^2 \frac{f_K^2
m_K^3 B_K}{3 \alpha^2 N_2^2}
\sum_{i=1}^3 \frac{1}{m_{h_i^0}^2}\left[ K_1 K_2 ({\cal U}_{2i}^2 -
{\cal U}_{3i}^2) - (K_1^2 - K_2^2)\, {\cal U}_{2i} {\cal U}_{3i} \right]
\label{imm12}
\end{equation}
The amplitude $M_{12}$ also receives contributions from the charged Higgs.
In the Appendix, we evaluate the relevant $H-W$ box diagram in Fig.\ 2,
showing that it can be neglected when compared with the values in
(\ref{dm0}) and (\ref{imm12}).

The neutral Higgs contribution to Im$A_0$ can be calculated from the tree
level diagram in Fig.\ 3(a). As we just want to get an approximate value,
it is reasonable to assume $(\mbox{Im}A_0)_{h^0} \simeq
(\mbox{Im}A_2)_{h^0}$. This means that the $\pi\pi$ isospin and charge
eigenstates give contributions of the same order. We use the $\pi^+\pi^-$
state, obtaining
\begin{eqnarray}
(\mbox{Im}A)_{h^0} & \simeq & - \Delta_{12}^d \frac{f_K m_\pi^2
m_K^2}{(m_u+m_d)} \frac{1}{\alpha^2 N_2^2} \sum_{i=1}^3
 \frac{1}{m_{h_i^0}^2} \left\{ m_d [ (1-\Delta_{11}^d) K_1 K_2
({\cal U}_{2i}^2 - {\cal U}_{3i}^2) \right.\nonumber \\
& & - K_2 K_3 {\cal U}_{2i}^2 + (K_2^2 (1-\Delta_{11}^d) +K_1(K_3
-K_1)) {\cal U}_{2i} {\cal U}_{3i} ] \nonumber \\
& & + \left.\left[ \begin{array}{c} m_d \leftrightarrow m_u \\
(1-\Delta_{11}^d) \leftrightarrow \Delta_{11}^u \end{array} \right]
- m_d K_1 \Delta_{11}^d {\cal U}_{2i} {\cal U}_{3i} \right\}
\label{imah0}
\end{eqnarray}

We also need to consider the $h^+$-mediated tree level diagrams shown in
Fig.\ 3(b). Repeating the above approximation we obtain
\begin{eqnarray}
(\mbox{Im}A)_{h^+} & = & \frac{1}{8}\frac{f_\pi m_\pi^2 m_K^2}{(m_u +m_d)}
\frac{1}{v_\eta v_\rho} \left(-\frac{v_s}{2\beta}\sin(2\gamma)\sin\varphi
\right)\frac{1}{m_{h^+}^2} \nonumber \\
& & \times [m_d (\sin \theta_c (V_L^d)_{11}(V_L^u)_{11}) -
\cos \theta_c (V_L^d)_{12}(V_L^u)_{11}) \nonumber \\
& & + m_u (\sin \theta_c (V_L^d)_{11}(V_L^u)_{11} +
\cos \theta_c (V_L^d)_{12}(V_L^u)_{11} - \sin\theta_c \cos\theta_c)]
\label{imah+}
\end{eqnarray}

Finally, to the same order of magnitude, it is necessary to evaluate the
``penguin-like'' diagram in Fig.\ 4. This conduces to
\begin{eqnarray}
(\mbox{Im}A_0)_{peng} & = & -\frac{1}{16\pi} \frac{m_c^2 m_s}{m_{h^+}^2}
\frac{1}{v_\eta v_\rho} \left(\ln (\frac{m_{h^+}^2}{m_c^2})-\frac{3}{2}
\right) \left(\frac{v_s}{2\beta} \sin (2\gamma) \sin \varphi\right)
\nonumber \\
& & \times (\sin \theta_c \cos \theta_c + \cos \theta_c (V_L^d)_{11}
(V_L^u)_{12} - \sin \theta_c (V_L^d)_{12}(V_L^u)_{12}) {\cal M}_{peng}
\label{imapeng}
\end{eqnarray}
where
\begin{equation}
{\cal M}_{peng} = \frac{g_s}{4\pi} \langle\pi\pi|i \bar s\sigma^{\mu\nu}
(1-\gamma_5)
\lambda^a dg_s\frac{q_\nu}{q^2} \bar q'\lambda^a \gamma_\mu q'|K^0\rangle
\end{equation}
(Notice that the penguin diagram contributes only to the $\Delta I=
\frac{1}{2}$ amplitude.)

\hfill

\noindent \underline{{\large Numerical analysis and conclusions}}

We have estimated the contributions to $\Delta m_K$, $\varepsilon$ and
$\varepsilon'$ coming from Higgs-mediated diagrams. Now, it is necessary
to check the compatibility with the experimental data.

The value of $\Delta m_K$ is accurately measured to be \cite{data}
\begin{equation}
\Delta m_K = (3.510\pm 0.018) \times 10^{-15} \mbox{GeV}
\label{dm}
\end{equation}
As this result agrees well with the Standard Model prediction, the Higgs
presence should not introduce significant modifications. In order to see
what happens, we refer to Eq.\ (\ref{dm0}), where we take $f_K=0.16$ GeV,
$m_K=0.5$ GeV and $\alpha^2\simeq 2\times 10^4\,\mbox{GeV}^2$ (see
relation (\ref{mw})). After doing this, the number of unknown parameters
is still very large. Nevertheless, in order to get an order of magnitude
for $(\Delta m_K)_{h^0}$, it is reasonable to assume that the ratio
$K_i/N_2$ and the mixing angles in ${\cal U}$ are approximate to 1. We
have then
\begin{equation}
(\Delta m_K)_{h^0} \sim (\Delta_{12}^d)^2 \frac{B_K}{m_{h_i^0}^2}
2\times 10^{-7} \mbox{GeV}^3
\label{dmkho}
\end{equation}
Using lattice calculations, the value of $B_K$ is found to be between 1
and $\frac{1}{3}$, while the neutral Higgs masses are uncertain. We will
take into account the theoretical bounds obtained for two-doublet models
\cite{babu}, hence we demand the value of $m_{h_i}$ not to be greater than
$\sim 150$ GeV. Now, if $(\Delta m_K)_{h^0}$ is imposed to be lower than
$(\Delta m_K)_{exp}$ at least by one order of magnitude we get the
constraint
\begin{equation}
(\Delta^d_{12}) = (V_L^d)_{11} (V_L^d)_{12} \alt 0.01
\label{cota}
\end{equation}
Notice that this constraint on $V_L^d$ has nothing to do with CP
violation, except for the fact that we have taken $\mu\approx v_\chi$.

It is important to mention that there is another contribution to $\Delta
m_K$ similar to that of Fig.\ 1. This is due to the presence of the
neutral gauge boson $Z'^0$, whose couplings with the ordinary quarks are
also not diagonal in flavor. As in the neutral scalar sector, the
experimental value of $\Delta m_K$ will also constrain the values of the
mixing angles in $V_L^d$ for a given value of $m_{Z'}$ \cite{tum}.

The next estimation to consider is that of $\varepsilon$. This parameter
is also accurately measured with the result
\begin{equation}
|\varepsilon|=(2.258\pm 0.018)\times 10^{-3}
\end{equation}

Let us consider the first term in (\ref{e}). Using (\ref{dm0}) and
(\ref{imm12}), we have
\begin{equation}
\frac{(\mbox{Im}M_{12})_{h^0}}{\sqrt{2}\Delta m_K} =
\frac{(\Delta m_K)_{h^0}}{4 \sqrt{2} \,(\Delta m_K)} {\cal F}
\end{equation}
where ${\cal F}$ represents the ratio between both sums in the expressions
of (Im$M_{12})_{h^0}$ and $(\Delta m_K)_{h^0}$.

The upper bound in (\ref{cota}) conduces to ${\cal F}\sim 0.1$. This does
not seem to be reasonable, since both sums in (\ref{dm0}) and
(\ref{imm12}) are in principle of the same order of
magnitude\footnote{This is not the case in the limit when the scalars
$h_i^0$ are degenerate in mass: due to the unitarity of ${\cal U}$, it is
immediate from (\ref{imm12}) that the contribution to $M_{12}$ vanishes.
Going to the next order in $(v_\eta/v_\chi)^2$, we could find a value for
$\varepsilon$ which is compatible with (\ref{cota}).}. On the contrary,
if we assume ${\cal F}\sim {\cal O}(1)$ we need for the angles in $V_L^d$
the stronger constraint
\begin{equation}
(V_L^d)_{11} (V_L^d)_{12} \alt 4\times 10^{-3}
\label{sevcota}
\end{equation}

The second term in (\ref{e}) can be treated using analogous approximations
as in the derivation of (\ref{dmkho}). From Eqs.\ (\ref{imah0}) and
(\ref{imah+}), our calculations lead to
\begin{mathletters}
\begin{eqnarray}
(\mbox{Im}A)_{h^0} & \sim & -\Delta_{12}^d \times 10^{-12} \mbox{GeV} \\
(\mbox{Im}A)_{h^+} & \sim & {\cal O}(10^{-14}) \mbox{GeV} \label{h+}
\end{eqnarray}\label{ima}
\end{mathletters}
These values have been obtained taking $m_{h^+}\sim m_{h^0}
\sim 150$ GeV together with angles $\gamma$ and $\varphi$ of order unity.
As is discussed above, there is no reason to think that the tree level
diagrams conducing to (\ref{ima}) produce a significant enhancement in one
of the amplitudes $A_I$. Hence, we take in both cases Im$A_0 \sim$ Im$A_2
\sim$ Im$A$.

The penguin diagram in Fig.\ 4 contributes solely to Im$A_0$. From
(\ref{imapeng}), the numerical estimation gives
\begin{equation}
\mbox{Im}A_0\sim {\cal O} (10^{-11} - 10^{-12}) \mbox{GeV}
\label{impen}
\end{equation}
where the uncertainty is fundamentally due to the mixing angles
combinations. The value of ${\cal M}_{peng}$ has been taken from bag model
calculations \cite{bag},
\begin{displaymath}
{\cal M}_{peng}= 2.0 \alpha_s \mbox{GeV}^2
\end{displaymath}
with $\alpha_s\sim 0.2$.

Now, taking into account the experimental value Re$A_0 \simeq 3.3\times
10^{-7}$ GeV, one can immediately see that both the tree level and penguin
contributions to $\varepsilon$ become vanishingly small. Thus, if we
demand the complex vacuum expectation values to be the sole CP violation
source in the theory, the experimentally measured value of $\varepsilon$
has to be given by the contribution from $(\mbox{Im}M_{12})_{h_i^0}$. That
is, the mixing angles in $V_L^d$ must reach the upper limit in
(\ref{sevcota}). The constraint is less severe in the limit when the three
``light'' neutral scalars have approximate masses.

Let us finally examine the $\varepsilon'$ parameter. Here, the
experimental data are not conclusive: different measurements give for
$\varepsilon'/\varepsilon$ \cite{data}
\begin{equation}
\mbox{Re}\left( \frac{\varepsilon'}{\varepsilon} \right) =
\left\{ \begin{array}{lc}
(2.3 \pm 0.65) \times 10^{-3} & \mbox{NA31} \\
(0.74 \pm 0.52 \pm 0.29) \times 10^{-3} \;\; & \mbox{E731}
\end{array} \right.
\label{epri}
\end{equation}
the latter being still compatible with the ``superweak'' mixing, where
$\varepsilon' = 0$. However, it is clear that both results in (\ref{epri})
imply an upper bound for $|\varepsilon'/\varepsilon|$ of order $10^{-3}$.

To see what happens in the $SU(3)_L\otimes U(1)_Y$ model, we refer to Eq.\
(\ref{e'}). Here, taking the above experimental values for Re$A_0$ and
$|\varepsilon|$, the ratio Re$A_2/$Re$A_0\simeq 1/22$ and the estimates in
(\ref{ima}) and (\ref{impen}), we have
\begin{equation}
\left| \frac{\varepsilon'}{\varepsilon} \right|_{SU(3)} \sim
{\cal O}(10^{-4})
\end{equation}
Thus, once we have fixed the value of $\varepsilon$, both $\Delta m_K$ and
$|\varepsilon'|$ fall very well within the experimental bounds.

It is interesting to compare our results with those obtained in Ref.\
\cite{lw} for a two-Higgs-doublet model. There, the authors introduced
arbitrary small parameters in the Lagrangian in order to reproduce the
observed CP violation and flavor changing phenomenological effects. The
required values for the parameters were tabulated in terms of the Higgs
masses and the VEV ratio $v_1/v_2$. In the $SU(3)_L \otimes U(1)_Y$ model,
the corresponding ``small'' parameter appears naturally as a mixing angle
in the matrix $V_L^d$, as shown in Eq.\ (\ref{sevcota}). Notice that
although both the $SU(3)_L\otimes U(1)_Y$ and the two-doublet models are
remarkably different, the result in Eq.\ (\ref{sevcota}) is comparable
with the bounds in Ref.\ \cite{lw}.

The reason for the small value of $V_{11}^d V_{12}^d$ remains hidden.
However, the observed hierarchies among the mixing angles in
$V_{CKM}=V_{L}^{u\dagger} V_{L}^d$ suggest the presence of new underlying
physics relating quark families, masses and mixing angles. If the $SU(3)_L
\otimes U(1)_Y$ were the theory beyond the Standard Model, the mechanism
responsible for a small angle compatible with (\ref{sevcota}) could
probably come to light within this context.

\acknowledgements

I would like to thank Dr.\ L.\ Epele for his helpful suggestions and a
critical revision of the manuscript. This work was partially supported by
CONICET (Argentina).

\pagebreak

\appendix
\section*{}

We estimate here the contribution to $\Delta m_K$ and $\varepsilon$ coming
from the box diagram shown in Fig.\ 2. The relevant quark-scalar vertices
in this graph are given by the interaction Lagrangian
\begin{eqnarray}
{\cal L} & = & \overline{D}\, \left[ \left(\frac{C_{22}^h}{v_\rho} \times
\openone^{(3\times 3)} + \left(\frac{C_{12}^h}{v_\eta} -
\frac{C_{22}^h}{v_\rho} \right) \, \Delta^d \right) V_{CKM}^\dagger\, \hat
M^u \frac{(1+\gamma_5)}{2} \right. \nonumber \\
& & \left. + \hat M^d\, V_{CKM}^\dagger \left(\frac{C_{12}^h}{v_\eta}
\times \openone^{(3\times 3)} - \left(\frac{C_{12}^h}{v_\eta} -
\frac{C_{22}^h}{v_\rho} \right) \, \Delta^u \right) \frac{(1-\gamma_5)}{2}
\right]\, U \, h^- + \mbox{ h.c.}
\end{eqnarray}
where $C^h$ represents the unitary matrix changing from the $(\eta_1^+, \rho^+,
h_1^+)$ basis to the mass eigenbasis $(G^+, h^+, H^+)$. In terms of the angles
$\gamma$ and $\phi$ defined in (\ref{higgmat}), we have
\begin{equation}
C_{12}^h = \frac{e^{i\theta_\eta}}{\alpha} \left( v_s \cos\gamma -
\frac{v_\eta v_\rho}{\beta} \sin\gamma\, e^{-i\varphi} \right)
\hspace{2cm}
C_{22}^h = - \frac{\alpha}{\beta} \sin\gamma\, e^{-i\varphi}
\end{equation}

The calculation of this type of diagrams has been performed in Refs.\
\cite{chang} and \cite{lw}, concerning the study of the Weinberg$-$Branco
model and the two-Higgs models respectively. We will not repeat the
details here. However, notice that in the $SU(3)_L \otimes U(1)_Y$ case
there is no discrete symmetry (exact or approximate) preventing the
presence of FCNI's. As a consequence, a contribution to the effective
Lagrangian which does not appear in Refs.\ \cite{chang} and \cite{lw} is
allowed. This new term is proportional to the Lorentz scalar operator
\begin{equation}
\bar s_L(q_1) \gamma^\alpha d_L(q_2)\, \bar s_L(p_1) \gamma_\alpha d_L(p_2)
\label{oper}
\end{equation}

Following the same steps as in Ref.\ \cite{chang} for the calculation of
the $K^0-\bar K^0$ amplitudes, we find
\begin{eqnarray}
(\mbox{Im}M_{12})_{box} & = & \frac{G_F}{\sqrt{2}}\,\frac{1}{12\pi^2}\,
\sin\theta_c \cos\theta_c\,\frac{m_K f_K^2 m_c^2}{m^2_{h^+}}\;
\frac{\mbox{Im}(C_{12}^h C_{22}^{h\ast})}{v_\eta v_\rho} \nonumber\\
& & \hspace{-2.3cm} \times\left[ \left( \ln (\frac{m_c^2}{m_W^2}) +
\frac{m_W^2}{m_h^2 - m_W^2} \ln (\frac{m_W^2}{m_h^2}) \right) m_c^2\,
(\sin\theta_c (V_L^d)_{12} (V_L^u)_{12} + \cos\theta_c (V_L^d)_{11}
(V_L^u)_{12}) \right. \nonumber \\
& & \hspace{-2.3cm} +\left. \frac{1}{8} m_K^2 \ln (\frac{m_c^2}{m_W^2}) \,
(1+ 3\frac{m_s^2}{m_K^2}) \, (\sin\theta_c (V_L^d)_{12} (V_L^u)_{12} -
\cos\theta_c (V_L^d)_{11} (V_L^u)_{12}-\sin\theta_c \cos\theta_c) \right]
\label{carg}
\end{eqnarray}

Except for very small values of the angles in $V_L^{d,u}$, we see that the
value of $(\mbox{Im}M_{12})_{box}$ is dominated by the first term in
(\ref{carg}). This is precisely the contribution due to the operator
(\ref{oper}). Now considering the constraint given by (\ref{sevcota}) and
all masses as in the evaluation of $(M_{12})_{h^0}$, we find that the
value of $(\mbox{Im}M_{12})_{box}$ can be at most one order of magnitude
smaller than the contribution of (\ref{imm12}). On the other hand, the
real part of $(M_{12})_{box}$ cannot be significantly different from the
imaginary part, thus the box contribution to $\Delta m_K$ may be safely
neglected.

\pagebreak

\begin{figure}[htbp]
\begin{center}
\epsfig{file=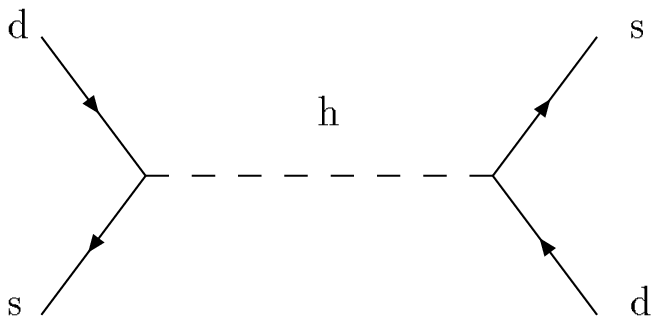}
\end{center}
\caption{Tree level contribution to $\Delta m_K$ and $\varepsilon$ due to
neutral Higgs boson exchanges.}
\end{figure}

\begin{figure}[htbp]
\begin{center}
\epsfig{file=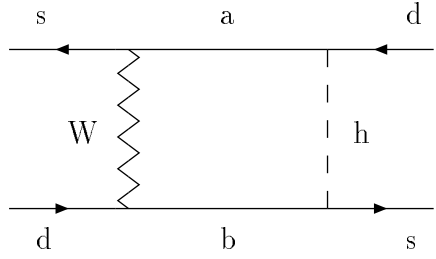}
\end{center}
\caption{Box diagram with one charged vector boson and one charged scalar
boson exchange.}
\end{figure}

\begin{figure}[htbp]
\begin{center}
\epsfig{file=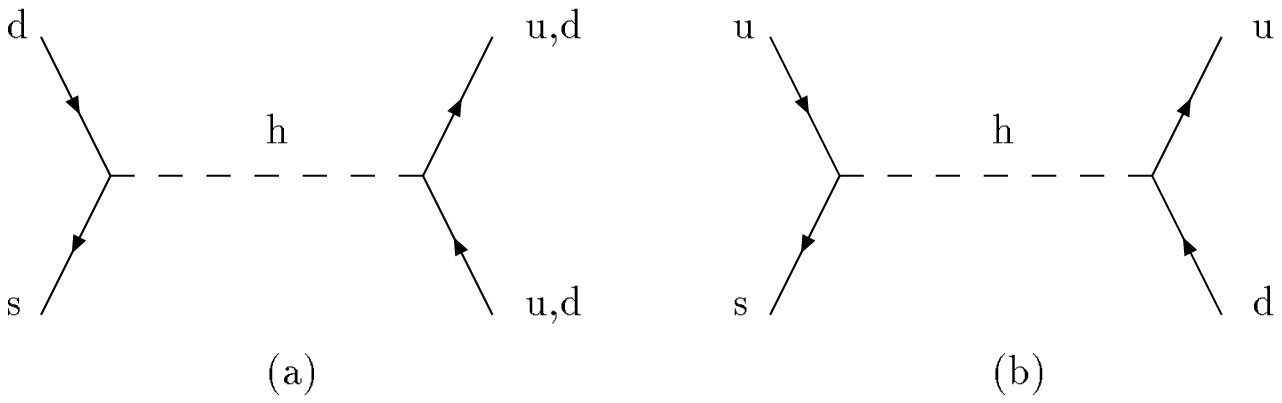}
\end{center}
\caption{(a) Tree level contribution to $\varepsilon'$ due to the exchange
of a neutral Higgs boson. \hfill \break (b) Tree level contribution to
$\varepsilon'$ due to the exchange of a charged Higgs boson.}
\end{figure}

\begin{figure}[htbp]
\begin{center}
\epsfig{file=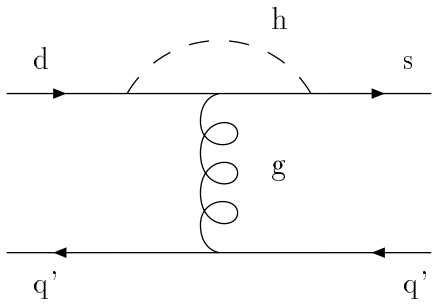}
\end{center}
\caption{Gluon penguin diagram mediated by a charged Higgs boson.}
\end{figure}

\end{document}